\documentclass[12pt]{article}
\usepackage{authblk}
\usepackage{graphicx}
\usepackage{amssymb}
\usepackage{amsmath}
\usepackage{mathrsfs}
\usepackage{boxedminipage}
\usepackage{epsfig}
\usepackage{color,soul}
\usepackage{enumerate}

\title{Analog dual to a 2+1-dimensional holographic superconductor}

\author[a,b]{Neven Bili\'c\thanks{bilic@irb.hr}}
\author[a]{J\'ulio C.\ Fabris\thanks{julio.fabris@cosmo-ufes.org}}
\affil[a]{Departamento de F\'isica, Universidade Federal do Esp\'irito Santo (UFES)
Av.\ Fernando Ferrari s/n CEP 29.075-910, Vit\'oria, ES, Brazil}
\affil[b]{Division of Theoretical Physics, Rudjer Bo\v skovi\'c Institute, 10002 Zagreb, Croatia}

\date{\today}
\begin{document}
\maketitle
\begin{abstract}
 We study an  analog hydrodynamic model that mimics 
 a 3+1 AdS planar BH spacetime dual to a 2+1-dimensional superconductor. 
 We demonstrate that the  AdS$_4$ bulk and its holographic dual
  could be realized in nature in  an analog gravity model based on fluid dynamics.
 In particular we mimic the  metric   
  of an $O_2$ holographic superconductor
and  calculate the entanglement entropy
 of a conveniently designed subsystem at the boundary
 of the analog AdS$_4$ bulk.
 
\end{abstract}

%


\section{Introduction}
A pseudo-Riemannian geometry of spacetime
can be mimicked by fluid dynamics in Minkowski spacetime. 
The basic idea is the emergence of an effective metric
\begin{equation}
G_{\mu\nu} = a [g_{\mu\nu}-(1-c_{\rm s}^2)u_\mu u_\nu],
\label{eq100}
\end{equation}
which describes the effective geometry for acoustic perturbations propagating in 
a fluid potential flow with  $u_\mu\propto \partial_\mu \theta$.
The quantity $c_{\rm s}$ is the adiabatic speed of sound, 
the conformal factor $a$ is 
related to the equation of state of the fluid, 
and the background spacetime metric $g_{\mu\nu}$ is usually assumed  Minkowski. 
The metric of the form (\ref{eq100})
has been exploited in various contexts
including 
emergent gravity \cite{babichev,novello2}, 
scalar theory of gravity \cite{novello}, 
Einstein-aether gravity \cite{jacobson},
acoustic geometry \cite{visser,bilic,kinoshita, barcelo2} and euclidean gravity \cite{barbero1,barbero2,mukohyama}.

The work presented here is motivated  by recent development of 
anti-de Sitter/conformal field theory
(AdS/CFT)  dual
theory of 2+1-dimensional  
superconductor 
\cite{hartnoll1,hartnoll2,gubser1,gauntlett,benini,horowitz1,aprile,bobev,albash,chakraborty}
(for a review and additional references see \cite{cai}).
The  AdS/CFT duality in these models is based on a correspondence  between gravitational theory
and dynamics of quantum field theory on the boundary of asymptotically anti-de Sitter (AdS) spacetime.
The gravity side can be well described by classical general relativity, while the dual field theory
involves the dynamics with strong interaction. This correspondence is often referred to as ``holography''
since a higher dimensional gravity system is described by a lower dimensional field theory
without gravity, which resembles optical holography.

 A particularly important work in this context is the minimal model of a holographic superconductor
by Bobev et al \cite{bobev} 
with an Abelian gauge field embedded in the truncation of four-dimensional maximal gauged super-gravity.
Besides, it is worth mentioning the work
on $d$-wave superconductivity by Benini et al \cite{benini} in which interesting physical phenomena are demonstrated
such as the formation of Fermi arcs.


The AdS$_4$ spacetime as a solution to Einstein's equations  cannot actually exist in nature due to instability problems. 
However, it can inspire some configurations where the underlying general gravitational structure can be studied through analogue models.
The aim of this paper is to demonstrate that AdS$_4$ and its holographic dual
 could be realized in nature in  an analog gravity model based on hydrodynamics
 of a physical fluid.
In particular we will mimic the bulk metric of  
the minimal model of a holographic superconductor
consisting of the metric, a charged scalar with a non-trivial potential and an Abelian
gauge field embedded in the truncation of four-dimensional maximal gauged super-gravity
\cite{bobev}.
This model was recently studied in the context of holographic entanglement entropy \cite{albash,chakraborty}.
	The entanglement entropy is an important tool  for keeping track of symmetry breaking and phase transition in strong coupling systems. In the context of black-hole thermodynamics the entropy of a black hole is proportional to the area of the horizon in the same way as is the entanglement entropy proportional to the boundary area of between two subsystems of a quantum system.

Our first  task is to derive an analog acoustic geometry  which mimics  a  $d+1$-dimensional asymptotic AdS geometry 
with a general planar black hole (BH). Furthermore, we will apply this to a 3+1-dimensional model
and calculate the entanglement entropy
for a particular geometry obtained as solution related to the holographic $O_2$ superconductor. 
The reason why we are specifically interested in 
the $O_2$ type is due to its pronounced first order 
phase transition at finite temperature.

 It is important to stress that analog gravity in general is concerned by curved geometry {\it per se}
 without referring to sources of the gravitational field
as  in general relativity. More specifically, the fluid analog mimics the geometry only and 
says nothing about the source such as matter and other fields. 
There are  no equations analog to Einstein's which, as in general relativity, would
involve curvature tensor and stress tensor.
However, even without Einstein's equations, the analog BH horizon entropy is realized via
quantum entanglement of phonons. This is why the model studied in this paper and analog gravity in general can teach us
something about black holes and related phenomena.

We divide the remainder of the paper into three sections and an two appendices.
We start with section  \ref{planar}
in which we derive an analog metric for a  $d+1$-dimensional AdS planar BH hole
of the form relevant for a holographic description of the superconductor.
In the next section, Sec.\ \ref{analog},
we apply our formalism to a 3+1-dimensional bulk related to 
the minimal model of the 2+1-dimensional holographic superconductor. For a particular geometry related to the $O_2$
superconductor we calculate the entanglement entropy.
Concluding remarks are given in section \ref{conclude}.
In appendix \ref{acoustic} we outline a derivation of the relativistic acoustic metric and in appendix \ref{external}
we derive the effective speed of sound in a fluid with an external pressure.


\section{Analog planar black hole}
\label{planar}

Geometric structures in the form of a planar BH may have interesting
 applications 
in condensed matter physics \cite{hartnoll}.
In this section we construct a model of an analog  planar BH hole in 
a general asymptotic AdS$_{d+1}$. 
A similar model for $d=4$ was discussed in detail 
 by Hossenfelder \cite{hossenfelder,hossenfelder2}
 and recently in \cite{nikolic,zingg}.
 We will discuss in more detail the case $d=3$ which is of particular interest
 for 2+1-dimensional superconductor \cite{bobev,albash,cai}. 
 In our approach we will consider a nonisentropic fluid flow
which yields the desired analog metric.

We start from a general form of the   AdS planar BH metric 
in an arbitrary number of space-like  dimensions $d$
\begin{eqnarray}
ds^2 = \frac{\ell^2}{z^2}  \left[e^{-\chi(z)}\gamma(z)   dt^2 - 
\gamma( z)^{-1} dz^2 -  d \mbox{\boldmath $x$}^2 \right],
\label{eq200}
\end{eqnarray}
where $\ell$ is the curvature radius of AdS$_{d+1}$ and
\begin{equation}
 d \mbox{\boldmath $x$}^2=  \sum_{i=1}^{d-1} {\rm d} x^i {\rm d} x^i .
 \label{eq204}
\end{equation}
For $d=3$ we will relate the functions
$\chi$ and $\gamma$  to the truncated
Lagrangian of the four-dimensional $\mathcal{N}=8$ super-gravity \cite{fischbacher}
studied by Bobev et al \cite{bobev} in the context of holographic superconductivity.
In order to have an asymptotic AdS for $z\rightarrow 0$ we can always rescale the time coordinate
so that, without loss of generality, we may assume 
\begin{equation}
\gamma(0)=1, \quad  \chi(0)=0.
\label{eq60}
\end{equation}
Next, the dimensionless functions $\chi$ and $\gamma$ can be thought of
as functions of the dimensionless variable $z/z_{\rm h}$,
 where $z=z_{\rm h}$
is the location of the horizon. In other words
\begin{equation}
\gamma(z_{\rm h})=0
\label{eq61}
\end{equation}
 and
$\gamma$ has no zeros on the interval $0<z<z_{\rm h}$.
Then, the horizon temperature is
 \begin{equation}
T=\left. \frac{e^{-\chi/2}}{4\pi}\frac{d\gamma}{dz}\right|_{z=z_{\rm h}} .
\label{eq2007}
\end{equation}
This temperature measured in some chosen fixed units, e.g., in units of $\ell^{-1}$ is ambiguous 
 because the geometry  (\ref{eq200}) 
is invariant under rescaling
 \begin{equation}
\tau\rightarrow \alpha\tau, \quad z\rightarrow \alpha z, 
\quad x^i\rightarrow \alpha x^i \quad z_{\rm h}\rightarrow \alpha{z}_{\rm h}. 
\label{eq2207}
\end{equation}
Thus, the metric (\ref{eq200}) has a  rescaled horizon $z_{\rm h}/\alpha$
with the corresponding rescaled horizon temperature
 \begin{equation}
\bar{T}=\left. \frac{e^{-\chi/2}}{4\pi}\frac{d\gamma(\alpha z)}{dz}\right|_{z=z_{\rm h}/\alpha}=\alpha T.
\label{eq2107}
\end{equation}
However, the temperature $T$ expressed in units of $1/z_{\rm h}$ is unique, i.e., the quantity $Tz_{\rm h}$ is invariant under the rescaling (\ref{eq2207}).
Therefore, in the following we will express the temperature and other dimensionfull physical quantities 
in units of some power of $z_{\rm h}$.

Now we seek a fluid analog model 
 which would mimic the induced metric of the form (\ref{eq200}).
The basic idea is to find a suitable coordinate transformation
$t\to \bar{t}$, $z\to \bar{z}$ such that the new metric 
takes the form of the relativistic acoustic metric (\ref{eq3008})
derived in appendix \ref{acoustic} with $g_{\mu\nu}$ replaced by the Minkowski metric $\eta_{\mu\nu}$
\begin{equation}
G_{\mu\nu}=\frac{n}{m^2 c_{\rm s} w}
[\eta_{\mu\nu}-(1-c_{\rm s}^2)u_\mu u_\nu]\, .
\label{eq3108}
\end{equation}
Here $n$ and $w$ denote the particle number density and specific enthalpy, respectively,
and an arbitrary mass scale $m$ is 
introduced to make $G_{\mu\nu}$ dimensionless. 
The specific enthalpy is defined as usual
\begin{equation}
w=\frac{p+\rho}{n} ,
\label{eq3018}
\end{equation}
where $p$ and $\rho$ denote the pressure and energy density, respectively.
The quantity $c_{\rm s}$ is the so-called ``adiabatic'' speed of sound
defined by
\begin{equation}
c_{\rm s}^{2}\equiv \left.\frac{\partial p}{\partial \rho} \right|_s =
\frac{n}{w}\left(\left. \frac{\partial n}{\partial w}\right|_s\right)^{-1} ,
\label{eq2011}
\end{equation}
where $|_s$ denotes that the specific entropy, i.e., entropy per particle $s=S/N$, is kept fixed.
The second equality in (\ref{eq2011}) follows from the thermodynamic law 
\begin{equation}
dw=Tds+\frac{1}{n} dp.
\label{eq2012}
\end{equation}

Following Hossenfelder \cite{hossenfelder} we
transform the metric (\ref{eq200})  by making use of 
a  coordinate transformation 
\begin{equation}
 t=\bar{t}+ h(z), \quad  z = z(\bar{z}) ,
 \label{eq120}
\end{equation}
where the functions $z(\bar{z})$ and $h(z)$ are determined by the requirement that
the transformed metric takes the form (\ref{eq3108}).
By simple algebraic manipulations the line element (\ref{eq200}) can be recast into 
a convenient form
\begin{eqnarray}
ds^2	&=&	\frac{\ell^2}{z^2} \biggl\{  d\bar{t}^2 - d\bar{z}^2 - d \mbox{\boldmath $x$}^2 
- (1-\tilde{\gamma}) d\bar{t}^2
\nonumber \\
	& &	
+ 2 (1-\tilde{\gamma})^{1/2}(c_{\rm s}^2-\tilde{\gamma})^{1/2}
	d\bar{t} d\bar{z} -(c_{\rm s}^2-\tilde{\gamma}) d\bar{z}^2 ] \biggr\},
\label{eq4001}
\end{eqnarray}
where we have set
\begin{equation}
\frac{dz}{d\bar{z}}  = e^{\chi/2} c_{\rm s} ,
\label{eq4002} 
\end{equation}
\begin{equation}
\frac{dh}{d z} = \frac{(1-\tilde{\gamma})^{1/2}
(c_{\rm s}^2-\tilde{\gamma})^{1/2}}{c_{\rm s} e^{\chi/2}\tilde{\gamma}} ,
\label{eq4003}
\end{equation}
and an abbreviation
\begin{equation}
\tilde{\gamma}=  e^{-\chi}\gamma .
\label{eq4000} 
\end{equation}
 From  (\ref{eq60}) and (\ref{eq61}) it follows
\begin{equation}
	0\leq \tilde{\gamma}\leq 1;  \quad \tilde{\gamma}(z_{\rm h})=0, \quad \tilde{\gamma}(0)=1 .
	\label{eq4200} 
\end{equation}
Comparing (\ref{eq4001}) with the acoustic metric (\ref{eq3108}) we identify 
$c_{\rm s}$ as
the speed of sound
and the non-vanishing components of the velocity vector
$u_{\bar{t}}$ and $u_{\bar{z}}$  in transformed coordinates as
\begin{equation}
u_{\bar{t}}= \frac{(1-\tilde{\gamma})^{1/2}}{(1-c_{\rm s}^2)^{1/2}}, \quad 
u_{\bar{z}}=- \frac{(c_{\rm s}^2-\tilde{\gamma})^{1/2}}{(1-c_{\rm s}^2)^{1/2}}.
 \label{eq123}
\end{equation}
These equations imply
\begin{equation}
\tilde{\gamma}\leq c_{\rm s}^2 \leq 1 .
\label{eq151}
\end{equation}
Next, by applying the potential-flow equation (see appendix \ref{acoustic})
\begin{equation}
w u_\mu =\partial_\mu\theta 
 \label{eq124}
\end{equation}
we derive closed expressions for $w$, $n$, and $c_{\rm s}$ 
in terms of the variable $z$. Since the metric is stationary, the velocity potential must be of the form
\begin{equation}
 \theta=m\bar{t} + g(z) ,
 \label{eq201}
\end{equation}
where $m$ is an arbitrary mass parameter which we can identify with the mass scale that appears 
 in (\ref{eq3108})
 and $g(z)$ is a function of $\bar{z}$ through $z$.
Then,  from (\ref{eq124}) and (\ref{eq201}) it follows 
\begin{equation}
 w=\frac{m}{u_{\bar{t}}}=m\frac{(1-c_{\rm s}^2)^{1/2}}{(1-\tilde{\gamma})^{1/2}},
 \label{eq101}
\end{equation}
and the function $g$ in (\ref{eq201}) must satisfy
\begin{equation}
\frac{dg}{dz}=w u_{\bar{z}}\left(\frac{dz}{d\bar{z}}\right)^{-1} 
=-\frac{m}{c_{\rm s}e^{\chi/2}}\frac{(c_{\rm s}^2-\tilde{\gamma})^{1/2}}{(1-\tilde{\gamma})^{1/2}}.
 \label{eq121}
\end{equation}
The particle number density can be obtained from the condition that
the conformal factor in (\ref{eq200}) must be equal to that of (\ref{eq3108}),  i.e., we require
\begin{equation}
\frac{n}{m^2 c_{\rm s} w}=\frac{\ell^2}{ z^2} .
\label{eq202}
\end{equation}
As $m$ is arbitrary it is natural to choose 
\begin{equation}
m=\frac{1}{\ell} ,
\label{eq203}
\end{equation}
so using this and (\ref{eq101}) we find
\begin{equation}
n= \frac{c_{\rm s}}{\ell z^2}\frac{(1-c_{\rm s}^2)^{1/2}}{(1-\tilde{\gamma})^{1/2}}.
 \label{eq4004}
\end{equation}
In this way, both $w$ and  $n$  are expressed as functions of $z$ and  $c_{\rm s}$.
However, $c_{\rm s}$ is not independent since by the definition (\ref{eq2011})
\begin{equation}
c_{\rm s}^{2}= 
\left.\frac{n}{w} \frac{\partial w}{\partial n}\right|_s =\frac{n}{w} 
\frac{dw}{dz}\left(\frac{dn}{dz}\right)^{-1}.
\label{eq3011}
\end{equation}
Using  (\ref{eq3011}) with (\ref{eq101}) and (\ref{eq4004}) we obtain a differential equation 
for $c_{\rm s}$
\begin{equation}
2 c_{\rm s} \frac{dc_{\rm s}}{dz}
-c_{\rm s}^2\left[\frac{2}{z} +\frac12 \frac{d}{dz}\ln(1-\tilde{\gamma})\right]
+\frac12 \frac{d}{dz}\ln(1-\tilde{\gamma})=0,
\label{eq4005}
\end{equation}
with solution
\begin{equation}
c_{\rm s}^2=1-\frac{z^2}{z_{\rm h}^2}(1-\tilde{\gamma})^{1/2}\left(K+2z_h^2\int_z^{z_{\rm h}}
 \frac{dz}{z^3}(1-\tilde{\gamma})^{-1/2}\right).
\label{eq106}
\end{equation}
The integration constant must satisfy the constraint $1\geq K\geq 0$  
as a consequence of the condition $0\leq c_{\rm s}^2\leq 1$.
Dimensionless physical quantities such as $\ell w$, $c_s$ 
and the components of the fluid velocity field
are functions of $z/z_{\rm h}$ and are invariant under the rescaling (\ref{eq2207}).

Plugging (\ref{eq106}) into (\ref{eq101}) and (\ref{eq202}) one obtains $w$ and $n$ as functions of $z$.
Note that explicit functional forms of  $z(\bar{z})$, $\gamma(z)$, and $g(z)$ can be obtained
by making use of (\ref{eq106}) and integrating respectively  (\ref{eq4002}), (\ref{eq4003}), and (\ref{eq121}).
However, the precise forms of these functions are not really needed for obtaining a closed expression
for the analog metric.

It is of particular interest to discuss the above solution in the asymptotic limit, i.e., in the limit 
$z\rightarrow 0$. 
Motivated by the asymptotic behavior of the $O_2$
holographic  superconductor with $d=3$
(see section \ref{superconduct})
\begin{eqnarray}
\tilde{\gamma}(z) =  1 + c_3 \left( \frac{ z}{z_{\rm h}}\right)^3 +\mathcal{O}(z^{3+1}),
\label{eq4108}
\end{eqnarray}
in the following  we assume for general $d$ 
\begin{eqnarray}
\tilde{\gamma}(z) =  1 + c_d \left( \frac{ z}{z_{\rm h}}\right)^d +\mathcal{O}(z^{d+1}),
\label{eq4008}
\end{eqnarray}
with $c_d<0$. 
Then, it may be easily shown that in the limit $z\rightarrow 0$ the sound speed squared tends to a constant 
$c_{\rm s}^2\rightarrow d/(d+4)<1$.
However, in this limit $\tilde{\gamma}\rightarrow 1$ so from equations (\ref{eq123}) it follows that the limit 
$z\rightarrow 0$ cannot be reached
since we must have $c_{\rm s}^2\geq \tilde{\gamma}$. 
This puts the constraint as to how close to the boundary is
our analog metric applicable. Our analog model breaks down at 
a point $z=z_{\rm min}$ which is
the maximal root of the equation $c_{\rm s}^2=\tilde{\gamma}$. 
For the minimal value of $K$, $K_{\rm min}=0$, this equation reads
\begin{equation}
 (1-\tilde{\gamma}(z))^{1/2}- 2z^2\int_z^{z_{\rm h}}
\frac{dy}{y^3}(1-\tilde{\gamma}(y))^{-1/2}=0 .
\label{eq119}
\end{equation}

In the case of a Schwarzschild AdS planar black hole, i.e., for  $\chi=0$ and $\gamma=1-(z/z_{\rm h})^d$, 
the integration in (\ref{eq4005}) 
can be easily performed yielding
\begin{equation}
c_{\rm s}^2=\frac{d}{d+4}+\left(\frac{4}{d+4}-K\right) \left(\frac{z}{z_{\rm h}}\right)^{d/2+2} ,
\label{eq4009}
\end{equation}
The condition $c_{\rm s}^2-\gamma=0$ now reads
\begin{equation}
\left(\frac{z}{z_{\rm h}}\right)^d +\left(\frac{4}{d+4}-K\right)\left(\frac{z}{z_{\rm h}}\right)^{d/2+2} - \frac{4}{d+4}=0 ,
\label{eq4010}
\end{equation}
For example, for $d=4$, the root $z_{\rm min}$ is given by
\begin{equation}
\frac{z_{\rm min}}{z_{\rm h}}= (3-2K)^{-1/4} \geq 3^{-1/4},
\label{eq4011}
\end{equation}
and for $d=3$ we find numerically
\begin{equation}
\frac{z_{\rm min}}{z_{\rm h}}= 0.727, \quad {\rm for}\quad K=K_{\rm min}=0.
\label{eq4111}
\end{equation}

Hence, the simple prescription  for an analog model is only valid from the point $z_{\rm min}$
up to the location of the horizon at $z_{\rm h}$.
In principle we could place the boundary of our model at $z_{\rm min}$ and cut off 
the section of AdS from $z=0$ to $z_{\rm min}$ as it has been done in the Randall-Sundrum model
\cite{randall1,randall2}. 
However, as we aim to make a connection with CFT at the boundary of AdS and calculate 
the boundary entanglement entropy at $z=0$, we would like to extend our model  all the way down to the AdS boundary at $z=0$.
As we demonstrate in appendix \ref{external}, such an an extension can be achieved by manipulating 
the equation of state by adding an external pressure.
For a fluid with an external pressure of the form
\begin{equation}
p_{\rm ext} = \alpha (p+\rho) ,
\label{eq6105}
\end{equation}
where $\alpha$ is a function of $z$, one finds the effective speed of sound
\begin{equation}
\tilde{c}_{\rm s}^{2}
=\frac{c_{\rm s}^2-\alpha}{1+\alpha} .
\label{eq6106}
\end{equation}
Depending on the functional form of $\tilde{\gamma}$
we can choose $\alpha$ to make the quantity 
$\tilde{c}_{\rm s}^2$ satisfy
equation
(\ref{eq151}) in the interval $0\leq z \leq z_{\rm h}$.
For example, if $\tilde{\gamma}$  behaves  as in
(\ref{eq4008}) near $z=0$,
we can choose
\begin{equation}
\alpha =\frac{d-(d+4)\tilde{\gamma}}{(d+4)(1+\tilde{\gamma)}}
\label{eq5008}
\end{equation}
to obtain $c_{\rm s}^2 \geq \tilde{\gamma}$ 
in the entire interval $0\leq z \leq z_{\rm h}$ and
\begin{equation}
\lim_{z\rightarrow 0} \tilde{c}_{\rm s}^2 =1.
\label{eq5013}
\end{equation}
 
\section{Analog bulk for the holographic superconductor}
\label{analog}
Here we consider a concrete example of the analog metric of the form (\ref{eq200})
for $d=3$ related to the holographic superconductor.
Instead of solving the field equations we will implement 
 the already known solutions \cite{bobev,albash,chakraborty} 
 into our analogue setup. Based on the known results we will construct
 approximate analytic expressions for $\gamma$ corresponding to a chosen horizon temperature.
With this we can calculate the entanglement entropy and by comparison with the
results of Refs. \cite{albash,chakraborty} we can also find an analytic expression for $\chi(z)$.
The analog geometry which  we have derived in
general form can be used to mimic these analytic expressions.

\subsection{Holographic superconductor}
\label{superconduct}
Here we briefly review the minimal model of a holographic superconductor
following Bobev et al \cite{bobev}. 
We consider 
the minimal model of a holographic superconductor  realized by 
an $SO(3)\times SO(3)$ invariant truncation of four-dimensional  $\mathcal{N}=8$ gauged super-gravity
\cite{fischbacher}.
The truncated action is 
\begin{equation}
S=\frac{1}{16 \pi G_4} \int d^4x \sqrt{-G} \left(- \mathcal{R}+\mathcal{L} \right),
\end{equation}
where  $\mathcal{L}$ involves 
two real dimensionless scalar  fields $\lambda$ and $\varphi$ coupled to an Abelian gauge field $A_\mu$
and gravity. The Lagrangian 
can be written as 
\begin{equation}
\mathcal{L} =    
- \frac{1}{4} F_{\mu \nu} F^{\mu \nu} + 
2 \partial_\mu \lambda \partial^\mu \lambda +\frac{\sinh^2 \left( 2 \lambda \right)}{2}
\left( \partial_\mu \varphi - \frac{g}{2} A_{\mu} \right) \left(  \partial^\mu \varphi  - 
\frac{g}{2} A^{\mu} \right) - \mathcal{P}  ,
\label{eq2001}
\end{equation}
with potential 
\begin{equation} \label{eqt:P}
\mathcal{P} = - g^2 \left( 6 \cosh^4 \lambda - 8 \cosh^2 \lambda \sinh^2 \lambda + 
\frac{3}{2} \sinh^4 \lambda \right) .
\end{equation}
The gauge coupling $g$ sets the scale $\ell$ of AdS$_4$ 
via the relation $\mathcal{P}\ell^2 = -6$ \cite{aprile}
with scalar potential evaluated at a critical point.
For the critical point $\lambda=0$ related to $SO(8)$ global symmetry \cite{bobev,fischbacher}
we obtain the relation $g^2\ell^2=1$.
The spacetime metric can be parameterized as 
\begin{equation}
d s^2 =\frac{\ell^2}{z^2}\left[ \gamma(z) e^{- \chi(z) } dt^2 -
\left( d x_1^2 + d x_2^2 \right)- \frac{d z^2}{\gamma(z)}\right] \ , 
\label{eq1000}
\end{equation}
where the functions $\gamma$ and $\chi$ are to be determined by solving the field equations
with appropriate boundary conditions.
As we have noted in section \ref{planar}, the value $\chi_0\equiv\chi(0)$ can be set to zero by rescaling the time 
coordinate.

The field equations  are derived in Ref.\ \cite{bobev}
for the gauge choice $\varphi=0$ and $A_\mu= (\psi(z),0,0,0)$ and solved for two types of superconductors
depending on the choice of boundary conditions, 
with non-trivial gauge fields and scalar condensates below some critical value of the temperature.
The solutions are characterized by the vacuum expectation values of the charged operators $O_1$ and $O_2$
(see Figs.\ 1 and 2 in Ref.\ \cite{bobev})).
Depending on the asymptotic behavior of the field $\lambda$ we distinguish two solutions:
\begin{description}
	\item[i)]
	$\lambda=\lambda_1 \tilde{z}+\mathcal{O}(\tilde{z}^3)$ 
	corresponding to an $O_1$ superconductor 
	with $O_1\propto \lambda_1$ and $O_2=0$, and 
	\item[ii)]
	$\lambda=\lambda_2 \tilde{z}^2+\mathcal{O}(\tilde{z}^4)$
	corresponding to an $O_2$ superconductor with
	$O_2\propto \lambda_2$  and $O_1=0$.
\end{description}
Here and from here on we use the dimensionless variable $\tilde{z}=z/\ell$.
As functions of temperature, the condensates $O_1$ and $O_2$ exhibit 
the second and first order phase transitions, respectively. The typical behavior of 
the condensates as functions of temperature is shown in 
Figs.\ 1 and 2 of  Ref.\ \cite{bobev}.
The quantity $\rho_{\rm c}$ which was chosen to set the units in these figures 
appears as a coefficient in the expansion $\psi=\mu\ell -\rho_{\rm c}\ell z +\dots$  near the AdS boundary.
Physically, $\mu$ and $\rho_{\rm c}$ are appropriately normalized chemical potential and charge density, respectively.
From the field equations one can derive
the following asymptotic expansions near $z=0$:
\begin{equation}
\lambda=\lambda_1\tilde{z}+ \lambda_2\tilde{z}^2
+\frac{\lambda_1}{24}\left(2\lambda_1^2-3 e^{\chi_0} \psi_0^2\right)\tilde{z}^3
+\mathcal{O}(\tilde{z}^4),
\label{eq2102}
\end{equation}
\begin{equation}
\psi=\psi_0+\psi_1\tilde{z}+ 
\frac{\psi_0}{2}\lambda_1^2 \tilde{z}^2+\frac{\psi_0}{3}\lambda_1\lambda_2 \tilde{z}^3
+\mathcal{O}(\tilde{z}^4),
\label{eq2104}
\end{equation}
\begin{equation}
\gamma=1+\lambda_1^2\tilde{z}^2 + \gamma_3\tilde{z}^3+ \mathcal{O}(\tilde{z}^4),
\label{eq2101}
\end{equation}
\begin{equation}
\chi=\chi_0+\lambda_1^2\tilde{z}^2 
+\frac83 \lambda_1\lambda_2 \tilde{z}^3
+\frac14\left(\lambda_1^4+ 8\lambda_2^2-e^{\chi_0}\lambda_1^2 \psi_0^2\right)\tilde{z}^4+ \mathcal{O}(\tilde{z}^5).
\label{eq2103}
\end{equation}
As we have mentioned, $\chi_0$ can be set to 0 and the other coefficients in the expansion
are related to physical quantities as follows: 
\begin{equation}
\lambda_1=4\ell O_1\,  ,  \quad \lambda_2=4\ell^2 O_2 \, ,
\label{eq3104}
\end{equation}
\begin{equation}
\psi_0=\ell\mu \, , \quad \psi_1=-\ell\rho_{\rm c}. 
\label{eq3105}
\end{equation}

For $\lambda=\chi=0$ there are no condensates and the solution is just the Reisner-Nordstrom (RN) AdS$_4$ 
planar BH
with
\begin{equation}
\gamma_{\rm RN} =1-(1+Q^2)\frac{z^3}{z_{\rm RN}^3} +Q^2\frac{z^4}{z_{\rm RN}^3}
\label{eq2000}
\end{equation}
and 
\begin{equation}
\psi_{\rm RN}=\frac{2Q\ell}{z_{\rm RN}}\left(1-\frac{z}{z_{\rm RN}}\right).
\label{eq2100}
\end{equation}
The charge squared $Q^2$ ranges  between 0 and 3  where $Q^2=0$ corresponds to a Schwarzschild AdS$_4$ planar BH and 
and $Q^2=3$ to the maximal RN AdS$_4$ planar BH.
%


%

\subsection{Entanglement entropy}
\label{entangle}
Here we present the calculation of the holographic entanglement entropy in the analogue model
discussed in section \ref{superconduct}.
%
Before we proceed to do that let us first discuss basic 
notions related to the entanglement entropy in general.

Supose we have a quantum system with the density of states 
matrix $\rho=|\Psi\rangle \langle \Psi|$.
If we divide the total system into two subsystems $A$ and $B$ 
we define the reduced density matrix for the subsystem $A$
 by taking a partial trace over the subsystem $B$. i.e.,
$\rho_{A}=\mathrm{tr}_{B}\, |\Psi\rangle \langle \Psi|$.
Then, the entanglement entropy defined as 
\begin{eqnarray}
	S_A = - \mathrm{tr}_{A}\, \rho_{A} \log \rho_{A}
\label{eq0}
\end{eqnarray}
is the entropy for an observer who can access information only from the subsystem $A$ and can receive no information from
$B$. The subsystem $B$ is analogous to the interior of
a black hole horizon for an observer outside of
the horizon. However, it is often not easy to compute the entanglement entropy, in particular in field theory in
3+1 or higher dimensions.

A convenient description of the entanglement entropy is derived in a $d+1$-dimensional field theory. It has been shown that the leading term of the entanglement entropy 
 can be expressed as the area law \cite{bombelli,srednicki}
\begin{equation}
S_A=a \frac{\mbox{Area}(\partial
	A)}{\ell^{d-1}}+\mbox{subleading terms},
\label{divarea}
\end{equation}
where $\partial A$ is the boundary of $A$, $\ell$ is an ultraviolet cutoff or the minimal length in the theory, and  $a$ is a constant which depends on the system. 
It is not accidental that this area law is of the same form as  the Bekenstein-Hawking entropy
 of black holes in 3+1 dimensions which is proportional to the area of the event horizon, 
 with the constants $d=3$, $a=1/4$, and $\ell$ equal to the Planck length.

	It is of particular relevance here that the 
	 entropy-area relation arises in the context of AdS/CFT duality. AdS/CFT, or gauge/gravity duality, is a correspondence between string theories in asymptotically anti-de Sitter bulk spacetimes and  certain conformal field theories living on the holograhic  boundary
	 \cite{maldacena}.
	 
	 According to AdS/CFT, the entanglement entropy being basically tied to the gravity in the bulk, should reflect fundamental features of the boundary gauge theory.
	 In this regard  we will study the so called 
	 {\em holographic entaglement entropy} in 3+1 dimensions
	 in the context of holographic superconductivity.
	 There is a subtle  difference between the usual entanglement entropy and holographic entanglement entropy: although both  obey the area low, 
	 in the case of holographic entanglement entropy, as we will shortly demonstrate, for a fixed two-dimensional  subsystem on the holographic boundary the area depends on the geometry in the bulk.
	   In particular, we expect that the holographic entanglement entropy in our model should exhibit the phase transition 
	   discussed in the previous section as demonstrated   by the temperature dependence  of the superconductor condensates \cite{bobev}.

The holographic entanglement entropy $S$ in 
a 2+1-dimensional boundary CFT for a subsystem $\mathcal{A}$ that has an
arbitrary one-dimensional boundary $\partial\mathcal{A}$ is defined 
by the following area law \cite{ryu,lewkowycz,engelhardt}
\begin{equation}
S=\frac{{\rm Area}(\Sigma)}{4\ell_{\rm Pl}^2} ,
\label{eq6000}
\end{equation}
where $\Sigma$ is the two-dimensional static minimal surface in
AdS$_4$ with boundary $\partial\mathcal{A}$
and $\ell_{\rm Pl}$ is the Planck length.

As we are dealing with an analog geometry
we will assume that there exist a minimal length,
typically of the order of the atomic separation,
below which the bulk description 
of the fluid fails. 
This length 
is referred to in the condensed matter literature as the coherence length,
where the meaning of the word "coherence" is different from that in optics.
Since it describes the distance over which the wave function of a BE condensate tends to its bulk
value when subjected to a localized perturbation, it is also referred to as the
healing length \cite{pethick}.
 In analog gravity systems,
a healing length $\ell_{\rm hl}$
plays the role of the Planck length \cite{uhlmann,girelli,fleurov,rinaldi,anderson}
and for a BE gas is typically of order $\ell_{\rm hl} \simeq 1/(mc_{\rm s})$ 
where $m$ is the boson mass. Hence, to calculate the entanglement entropy we use
(\ref{eq6000}) with the Planck length $\ell_{\rm Pl}$ replaced by the healing length $\ell_{\rm hl}$. 
Furthermore, we will identify the arbitrary scale $\ell$ with $\ell_{\rm hl}$.
\begin{figure}[t]
\begin{center}
\includegraphics[width=0.8\textwidth,trim= 0 0cm 0 0cm]{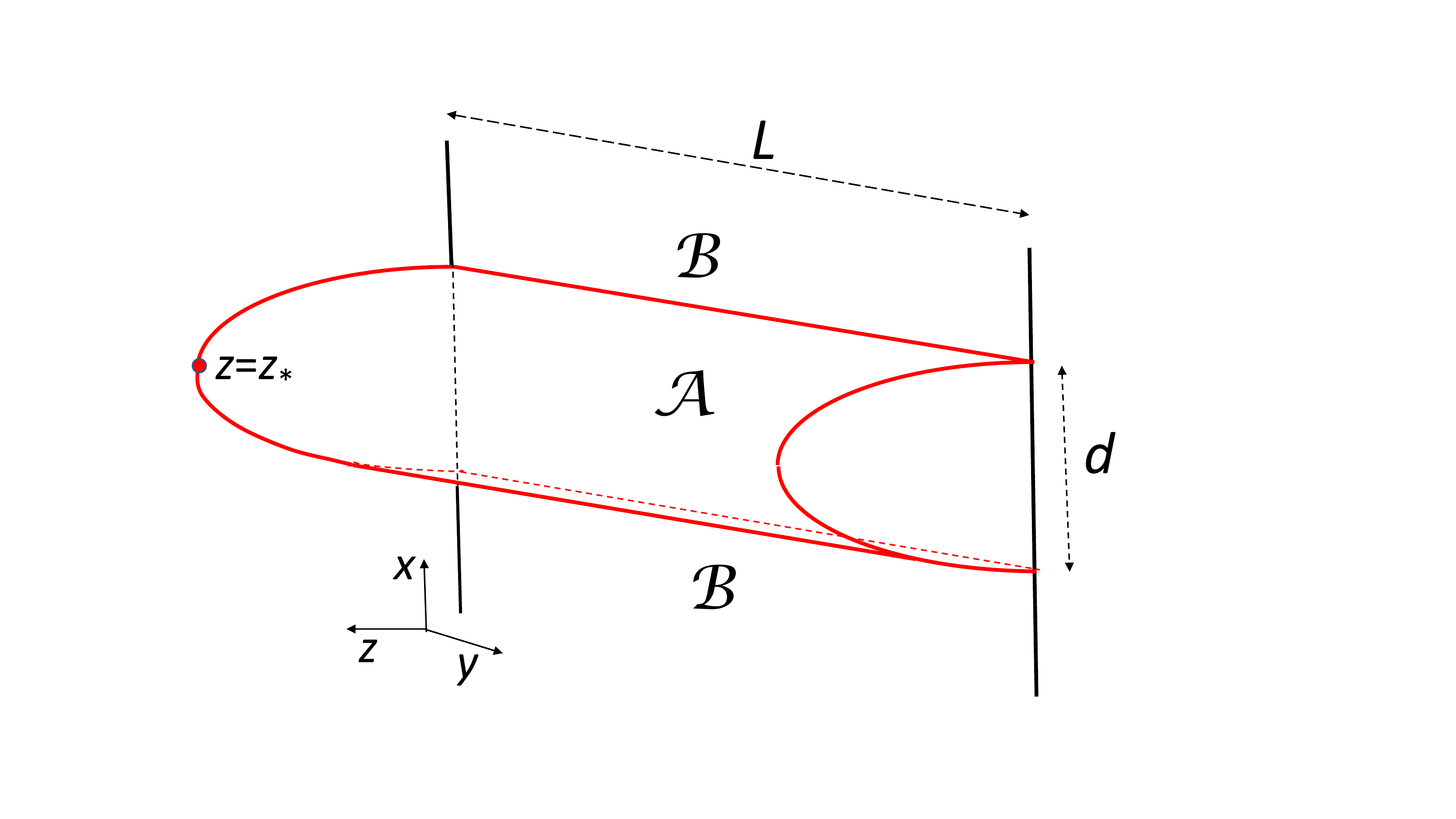}
\caption{Strip geometry employed 
to calculate the entanglement entropy. 
}
\label{fig2}
\end{center}
\end{figure}
Next we apply the prescription (\ref{eq6000}) to the geometry suggested in Refs.\ \cite{albash,ryu} 
illustrated in Fig.\ \ref{fig2} and calculate the entropy $S$ as a function of the strip width $d$ for
a fixed temperature. 

Consider the bulk metric (\ref{eq200}) with $d=3$ and
a surface $\Sigma$
defined by the equation
\begin{equation}
z-z(x)=0 ,
\label{eq6001}
\end{equation}
where $z(x)$ is a function of $x$ such that $\Sigma$ extends into the bulk and is 
bounded by the perimeter of $\mathcal{A}$  as illustrated in Fig.\ \ref{fig2}.
The induced metric $\sigma_{ij}$ on $\Sigma$ defines   
the line element
\begin{equation}
ds_\Sigma^2=\sigma_{ij}dx^idx^j= \frac{\ell^2}{z^2}\left[dx^2\left(1+\frac{{z'}^2}{\gamma}\right)+y^2\right].
\label{eq6002}
\end{equation}
The area of $\Sigma$ can be viewed as a functional 
\begin{equation}
	I[z,z']=-{\rm Area}(\Sigma)/L= \int_{-d/2}^{d/2}dx \mathcal{L},
	\label{eq6003}
\end{equation}
where $L$ and $d$ are respectively the length and width of the strip, and 
\begin{equation}
	\mathcal{L}= -\frac{\ell^2}{z^2}\left(1+\frac{{z'}^2}{\gamma}\right)^{1/2}.
	\label{eq00}
\end{equation}
Next we calculate  the maximal area of $\Sigma$.
Clearly, a maximum of Area  corresponds to a minimum of $I$ and variation of $I$ yields the equation of motion for $z$.
Instead of solving the equation of motion we will use  
the Hamiltonian approach. 
We define the conjugate momentum 
\begin{equation}
	\pi=\frac{\partial\mathcal{L}}{\partial z'} 
	\label{eq1}
\end{equation}
and construct the Hamiltonian
\begin{equation}
	\mathcal{H}=\pi z' - \mathcal{L} = \frac{\ell^2}{z^2}\frac{1}{(1+{z'}^2/\gamma)^{1/2}}.
	\label{eq2}
\end{equation}
It can be easily shown that the equation of motion is satisfied if and only if the Hamiltonian is a constant of motion. In particular, at the bottom of the surface $z=z_*$ we have $z'=0$ and the Hamiltonian 
is equal to $\ell^2/z_*^2$. In this way we obtain the equation
\begin{equation}
	\frac{\ell^2}{z_*^2}=	\frac{\ell^2}{z^2}\frac{1}{(1+{z'}^2/\gamma)^{1/2}},
	\label{eq4}
\end{equation}
from which we can  express $z'$ as
\begin{equation}
	z'=\pm	\frac{\sqrt{(z_*^4-z^4)\gamma}}{z^2} .	
	\label{eq55}
\end{equation}
Inserting this into (\ref{eq6003}) 
and changing the integration variable from $x$ to $z$ with $dx=dz/z'$ we obtain the area of the extremal surface
\begin{equation}
	{\rm Area} =8L\int_0^{z_*}dz\frac{z_*^2}{z^2}\frac{\ell^2}{\sqrt{(z_*^4-z^4)\gamma}}.
	\label{eq6}
\end{equation}
Dividing this by $4\ell^2$ we obtain 
the entanglement entropy 
expressed as an integral over $z$
\begin{equation}
S=\frac{{\rm Area}}{4\ell^2}=2L\int_0^{z_*}dz\frac{z_*^2}{z^2}\frac{1}{\sqrt{(z_*^4-z^4)\gamma}}.
\label{eq6004}
\end{equation}
The location of the bottom $z_*$  of the extremal surface is related to the strip width
\begin{equation}
d=2 \int_{-d/2}^{d/2}dx=2\int_0^{z_*}dz\frac{z^2}{\sqrt{(z_*^4-z^4)\gamma}}.
\label{eq6005}
\end{equation}

The integral in (\ref{eq6004}) is divergent near $z=0$ and can be regularized 
by adding and subtracting  a counter-term 
\begin{equation}
2L\int_\epsilon^{z_*} dz/z^2.
\label{eq6006}
\end{equation}
The entropy is then expressed as
\begin{equation}
S=S_{\rm fin}+\frac{2L}{\epsilon} ,
\label{eq6007}
\end{equation}
where the finite part reads
\begin{equation}
S_{\rm fin}=2L\int_0^{z_*}dz\left(\frac{z_*^2}{z^2}\frac{1}{\sqrt{(z_*^4-z^4)\gamma}}-\frac{1}{z^2}\right)-\frac{2L}{z_*} .
\label{eq6008}
\end{equation}

\begin{figure}[t]
\begin{center}
\includegraphics[width=0.8\textwidth,trim= 0 0cm 0 0cm]{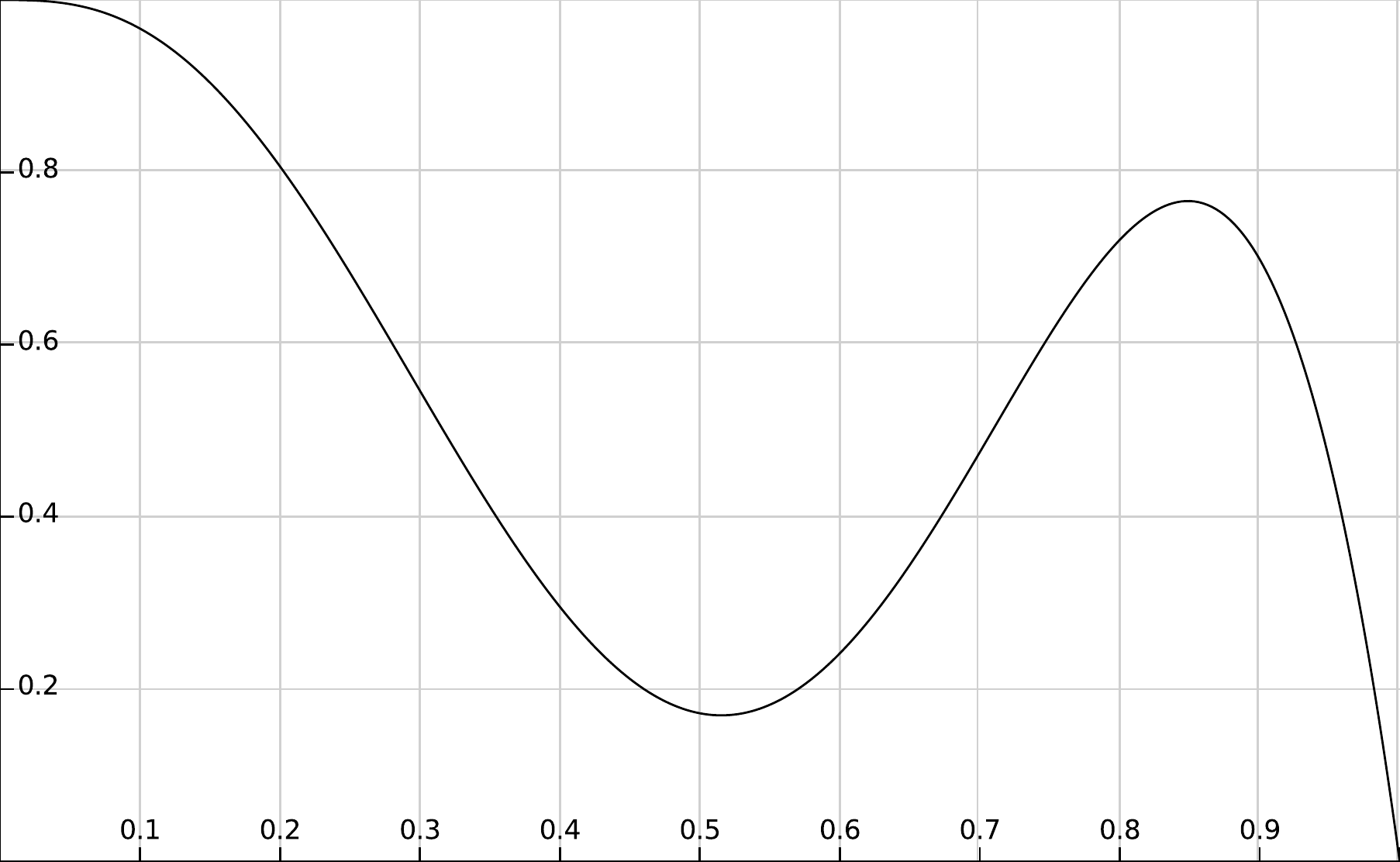}

\caption{The metric function $\gamma$ versus $z/z_{\rm h}$ 
 for $T=0.61 \times 10^{-2}\sqrt{\rho_{\rm c}}$.  
}
\label{fig3}
\end{center}
\end{figure}

Next we calculate the entanglement entropy using the bulk profile corresponding to an $O_2$ superconductor
at fixed temperature. The reason why we specifically
address the  $O_2$ type is that the
$O_2$ superconductor exhibits a first order 
phase transition which manifests itself as a discontinuity depicted in 
Fig.\ 2 of Ref.\ \cite{bobev}. 
To calculate $S_{\rm fin}$ 
 we use a polynomial function
\begin{equation}
\gamma=1+\sum_{i=3}^6c_i \left(\frac{z}{z_{\rm h}}\right)^i
\label{eq2105}
\end{equation}
with
\begin{equation}
c_3=-44, \quad c_4=118,\quad c_5=-98,\quad c_6=23.
\label{eq2205}
\end{equation}
We plot this function in Fig.\ \ref{fig3}.
This choice is motivated by 
the superconductor bulk metric profile
plotted in Fig.\ 7(b) of Ref.\ \cite{chakraborty}
for a fixed horizon temperature $T=0.61 \times 10^{-2}\sqrt{\rho_{\rm c}}$ 
where $\rho_{\rm c}$ is the charge density of the $O_2$ superconductor
(see section \ref{superconduct}). 
The function (\ref{eq2105}) is an analytic approximation to the bulk metric
found by numerically solving the field equations of the holographic superconductor.

\begin{figure}[t!]
\begin{center}
\includegraphics[width=0.45\textwidth,trim= 0 0cm 0 0cm]{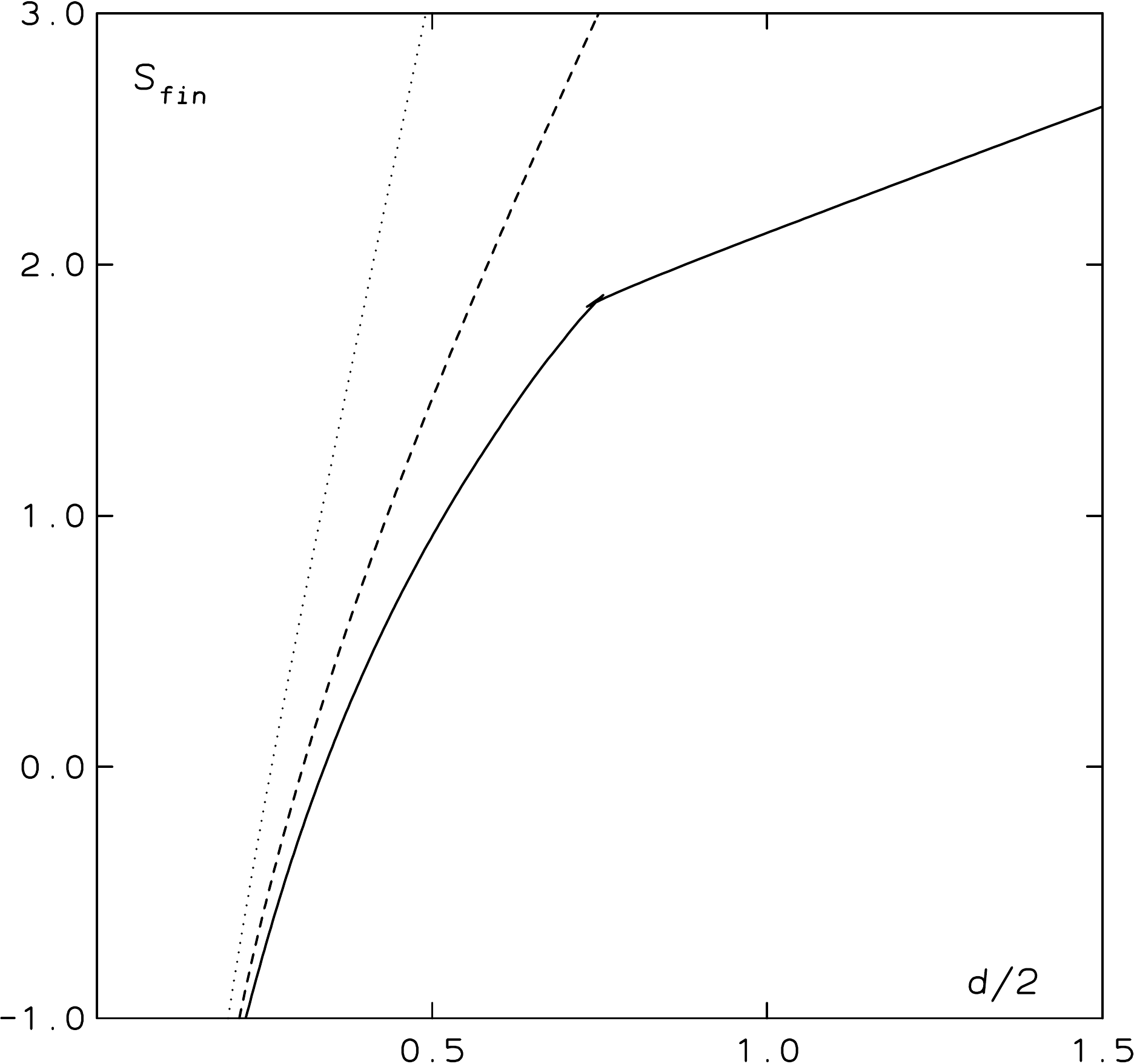}
\hspace{0.03\textwidth}
\includegraphics[width=0.45\textwidth,trim= 0 0cm 0 0cm]{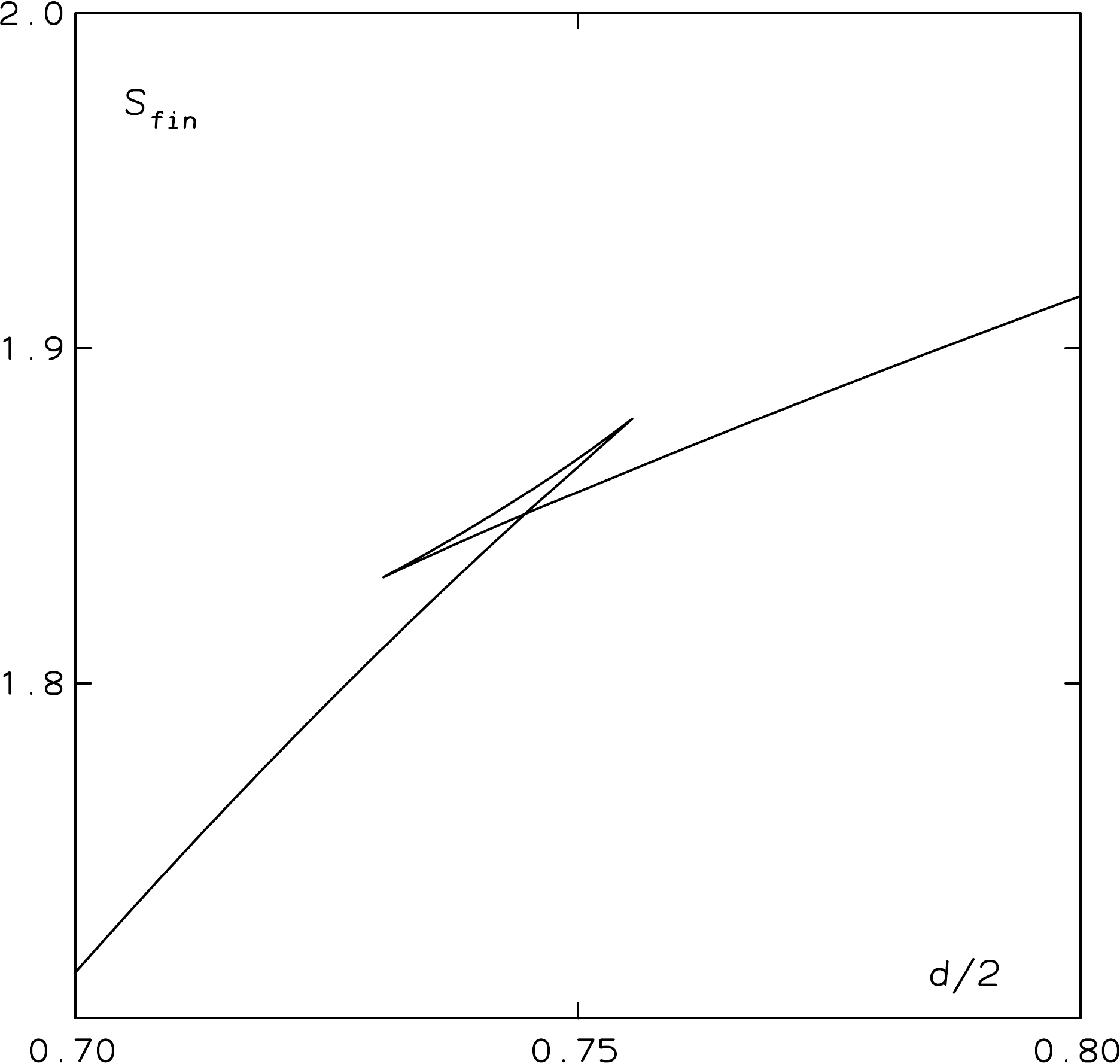}
\caption{The finite part of the entanglement entropy $S_{\rm fin}$ in units of $2L/z_{\rm h}$ versus half strip width $d/2$ in units 
of $z_{\rm h}$ at fixed temperature. Dotted and dashed lines represent the entanglement entropies of the Schwarzschild 
AdS planar and maximal RN AdS planar BH, respectively. 
The right panel shows the zoomed-in crossover region.}
\label{fig4}
\end{center}
\end{figure}

 In Fig.\ \ref{fig4} we plot $S_{\rm fin}$ as a function of $d/2$. For comparison we plot in the same figure the 
entanglement entropies of a Schwarzschild  AdS planar BH hole and a maximal  RN AdS planar BH 
which have the same asymptotic behavior near $z=0$. The metric profiles are determined so that the cubic terms are
the same as in the $O_2$ superconductor case. Hence we have
\begin{equation}
\gamma_{\rm AdS}=1+c_3\left(\frac{z}{z_{\rm h}}\right)^3
\label{eq5105}
\end{equation}
for the Schwarzschild AdS planar BH and
\begin{equation}
\gamma_{\rm RN}=1+c_3\left(\frac{z}{z_{\rm h}}\right)^3+3 \left(\frac{c_3}{4}\right)^{4/3} \left(\frac{z}{z_{\rm h}}\right)^4
\label{eq5106}
\end{equation}
for the maximal RN AdS planar BH.
The coefficient of the quartic term in (\ref{eq5106}) was fixed by virtue of (\ref{eq2000})
 and requirement 
$\gamma_{\rm RN}(z_{\rm RN}) =0$, where $z_{\rm RN}= (-4/c_3)^{1/3}z_{\rm h}$ is the location
of the RN BH horizon.

To complete our model we still have to determine the function $\chi(z)$.
To do this we need to set the scale $z_{\rm h}$ in relation to the previous works \cite{bobev,albash,chakraborty}.
We will make a comparison of the scales at a fixed temperature $T=0.61 \times 10^{-2}\sqrt{\rho_{\rm c}}$,
where  $\rho_{\rm c}$ is the charge density of dimension of length$^{-2}$.
In  Ref.\ \cite{bobev} $\rho_{\rm c}$ is chosen to set the scale
whereas in Refs.\ \cite{albash,chakraborty} the scale is set by the quantity
\begin{equation}
\tilde{\rho}_{\rm c}=\frac{\rho_{\rm c}\sqrt{16\pi G_4}}{\ell} .
\label{eq5205}
\end{equation}
The relation between $\tilde{\rho}_{\rm c}$ and $\rho_{\rm c}$ can be fixed by 
identifying the $O_2$ phase transition temperature $T_{\rm tr}$  of 
Albash and Johnson \cite{albash} (their figure 2(b))
$T_{\rm tr}=0.003635\tilde{\rho}_{\rm c}^{1/2}$
with that of
Bobev et al \cite{bobev} (their figure 2)
$T_{\rm tr}=0.007269\rho_{\rm c}^{1/2}$.
From this we obtain
\begin{equation}
\tilde{\rho}_{\rm c}=4\rho_{\rm c}.
\label{eq5203}
\end{equation}
In our approach the scale is set by $z_{\rm h}$ so we have to find a relation between our
 $z_{\rm h}$ and $\tilde{\rho}_{\rm c}$ or $\rho_{\rm c}$.
To this end we compare the transition point $d_{\rm tr}/2=0.744 z_{\rm h}$ (Fig.\ \ref{fig4})
with that of  Chakraborty \cite{chakraborty} $d_{\rm tr}/2=2.56\tilde{\rho}_{\rm c}^{-1/2}$.
This yields
\begin{equation}
\frac{1}{z_{\rm h}}=0.2906\tilde{\rho}_{\rm c}^{1/2}=0.5812\rho_{\rm c}^{1/2}.
\label{eq5204}
\end{equation}
Using this we can express the horizon temperature of our configuration depicted in Fig. \ref{fig3}
in units of $z_h^{-1}$,
 \begin{equation}
T z_{\rm h}\equiv\frac{3}{\pi} e^{-\chi_{\rm h}/2}=1.05\times 10^{-2} ,
\label{eq5206}
\end{equation}
which yields 
\begin{equation}
\chi_{\rm h} \equiv\chi(z_{\rm h})=-2\ln \frac{0.0105 \pi}{3}=9.02 .
\label{eq5207}
\end{equation}
Next, we express $\chi$ as a function of $z$ using the expression (\ref{eq2103}) from section \ref{superconduct}
in which we set $\chi_0=0$, 
 $\lambda_1=0$,  keep the $z^5$ term and neglect the higher order terms. Hence we write
\begin{equation}
\chi(z)=2\lambda_2^2\frac{z^4}{\ell^4}+\chi_5 \frac{z^5}{\ell^5} ,
\label{eq5208}
\end{equation}
where the coefficient
$\lambda_2$ 
can be fixed from Eq.\ (\ref{eq3104}) with 
the value of $O_2$  deduced from Fig.\ 2 of Ref.\ \cite{bobev}. 
At $T=0.61\times 10^{-2}\sqrt{\rho_{\rm c}}$
we find $\lambda_2=1.1462$ and using (\ref{eq5207}) we obtain
\begin{equation}
\chi(z)=2.63\left(\frac{z}{z_{\rm h}}\right)^4+6.39\left(\frac{z}{z_{\rm h}}\right)^5 .
\label{eq5209}
\end{equation}
This equation  together with  (\ref{eq2105}) and (\ref{eq2205}) 
can be used to find closed expressions for the hydrodynamic functions and variables of our
analog model.

The considerations in this section
 can as well be carried out for the type $O_1$ superconductor.

\section{Summary and conclusions}
\label{conclude}
We have derived an analog acoustic geometry  which mimics  a  $d+1$-dimensional asymptotic AdS geometry 
with a planar Black hole. In 3+1 dimensions, this geometry 
 has been exploited as a holographic model for the 2+1-dimensional superconductor.
We have applied this general analog geometry to a 3+1-dimensional bulk
 and calculated the entanglement entropy
 for a particular geometry obtained as solution related to the holographic $O_2$ superconductor. 
We have demonstrated that the entanglement entropy in our analog model 
exhibits the usual first order phase transition which characterizes the $O_2$ superconductor.

In this way we have confirmed the basic idea that a 3+1 AdS bulk with a planar BH  
can  be realized in nature as a  hydrodynamic analog gravity model. Moreover, the analog bulk metric can be parameterized so that the coefficient in the asymptotic expansion in powers of $z$ are such that the dual AdS/CFT boundary field theory corresponds to  the type $O_2$ superconductor. A procedure similar to the one described in 
section \ref{entangle} can easily be applied to the case of type $O_1$ superconductor.

 It would be of considerable interest to construct  a concrete fluid system in the laboratory which would satisfy the properties of the analog geometry described above.	
	 It is fare to say that at this stage we cannot provide a  clear proposal of how to prepare an adequate  laboratory setup.
As we are concerned with fluid velocities close to the speed of light $c$ and sound speed close to $c$,
we would need an essentially relativistic fluid.
So far the only  known realistic experimental set up for a relativistic-fluid laboratory is provided by high-energy
colliders. The study of analog gravity
in high energy collisions may in general improve our
understanding of the dynamics of general relativistic fluids
\cite{bilic2,bilic3}.
Maybe, with the advance of accelerator technology,
one day it will be possible, e.g., by choosing appropriate
heavy ions and specially designed beam geometry to
obtain the desired equation of state and expansion flow of
the fluid.

\section*{Acknowledgments}
 The work of N.~Bili\'c  has been partially supported by
the European Union through the European Regional Development Fund - the Competitiveness and
Cohesion Operational Programme (KK.01.1.1.06).
J.C.~Fabris thanks  CNPq (Brazil) and FAPES (Brazil) for partial support.

\appendix

\section{Acoustic metric}
\label{acoustic}

Here we briefly review the derivation of the relativistic acoustic metric.
Acoustic metric is the effective metric 
 perceived by acoustic perturbations propagating in a perfect fluid
background. 
Under certain conditions
the perturbations satisfy a Klein-Gordon equation in curved geometry 
with metric of the form (\ref{eq100}).

We first derive a propagation equation for linear perturbations 
of a nonisentropic flow assuming a fixed background geometry.
Following Landau and Lifshitz \cite{landau}  we assume that the enthalpy flow 
$w u_{\mu}$ is a gradient of a scalar potential, i.e., that there exist a scalar function $\theta$ such that
the velocity field satisfies 
\begin{equation}
w u_\mu =\partial_\mu\theta ,
 \label{eq403}
\end{equation}
where $w$ is the specific enthalpy defined by (\ref{eq3018}).
Then,
from the relativistic Euler equation and standard thermodynamic  identities
it follows \cite{nikolic} that the entropy gradient
is also proportional to the gradient of the potential, i.e.,
\begin{equation}\label{rnif10}
s_{,\mu}=w^{-1} u^{\nu}s_{,\nu}\theta_{,\mu} .
\end{equation}
Furthermore, instead of the continuity equation $(nu^{\mu})_{;\mu}=0$, one finds 
\begin{equation}
(nu^{\mu})_{;\mu}=\frac{1}{w}\frac{\partial p}{\partial s}u^{\mu}s_{,\mu} .
\label{eq442}
\end{equation}
In a nonisentropic flow we 
have $u^\mu s_{,\mu} \neq 0$ and the above equation shows that the particle number 
is generally not conserved.
As demonstrated in Ref.\ \cite{nikolic}, from equation (\ref{rnif10}) and Lagrangian description of fluid dynamics
it follows that the specific entropy is a function of the velocity potential $\theta$ only.
Then, using (\ref{rnif10}) equation (\ref{eq442}) can be expressed in the form
\begin{equation}
(n u^\mu)_{;\mu}
=\frac{\partial p}{\partial\theta},
\label{eq439}
\end{equation}
where $p=p(w,s(\theta))$ is the pressure of the fluid.

Given  some average bulk motion represented by
 $w$, $n$, and
 $u^{\mu}$, following the standard procedure \cite{visser,bilic,landau}, 
we make a replacement
 \begin{equation}
w\rightarrow w+\delta w, \quad n\rightarrow n+\delta n ,
\quad
u^{\mu}\rightarrow u^{\mu}+\delta u^{\mu},
\label{eq008}
\end{equation}
where the perturbations 
$\delta w$, 
$\delta n$, and 
$\delta u^{\mu}$ 
are induced by a small perturbation $\delta\theta$ 
around a background velocity potential $\theta$.
From (\ref{eq403}) 
it follows
\begin{equation}
\delta w=u^\mu\delta\theta_{,\mu},
 \label{eq406}
\end{equation}
\begin{equation}
w\delta u^\mu=(g^{\mu\nu}-u^\mu u^\nu)\delta\theta_{,\nu}.
 \label{eq407}
\end{equation}
Using this and (\ref{eq008})
equation
(\ref{eq439})
at linear order yields
\begin{equation}
\left(f^{\mu\nu}
\delta\theta_{,\nu} \right)_{;\mu}
+\left[\left( 
\frac{\partial n}{\partial\theta}u^\mu\right)_{;\mu}
-\left(\frac{\partial^2 p}{\partial\theta^2}\right)\right]\delta\theta=0,
 \label{eq413}
\end{equation}
where 
\begin{equation}
f^{\mu\nu}=\frac{n}{w}\left[g^{\mu\nu} -\left(1-\frac{w}{n}\frac{\partial n}{\partial w}\right)u^\mu u^\nu\right] .
 \label{eq423}
\end{equation}
Then, it may be easily shown
that equation (\ref{eq413}) can be recast into the form 
\begin{equation}
\frac{1}{\sqrt{-G}}
\partial_{\mu}
\left(
{\sqrt{-G}}\,G^{\mu\nu}
 \partial_{\nu}\delta\theta\right) + m_{\rm eff}^2 \delta\theta
=0 ,
\label{eq5}
\end{equation}
where the matrix $G^{\mu\nu}$
is the inverse of
the acoustic metric tensor 
\begin{equation}
G_{\mu\nu}=\frac{n}{m^2 c_{\rm s} w}
[g_{\mu\nu}-(1-c_{\rm s}^2)u_\mu u_\nu]\, ,
\label{eq3008}
\end{equation}
with determinant $G$.
Here $m$ is an arbitrary mass parameter
introduced to make $G_{\mu\nu}$ dimensionless and 
$c_{\rm s}$ is the speed of sound
defined by (\ref{eq2011}).

The effective mass squared is given by 
\begin{equation}
m^2 \sqrt{|G|}\, m_{\rm eff}^2= 
\left[\left(
\frac{\partial n}{\partial \theta}u^\mu\right)_{;\mu}
-\frac{\partial^2 p}{\partial \theta^2}\right] .
 \label{eq420}
\end{equation}
Hence, the linear perturbations $\chi$ propagate in the effective metric 
(\ref{eq3008})
and acquire an effective mass.

In an equivalent field-theoretical description \cite{babichev,nikolic,piattella} the fluid velocity $u_\mu$ is derived from the scalar field as 
$u_\mu= \partial_\mu \theta/\sqrt{X}$,  and $n$ and $c_{\rm s}$ are expressed in terms of the Lagrangian and its first 
and second derivatives with respect to the kinetic energy term $X=g^{\mu\nu}\theta_{,\mu}\theta_{,\nu}$.
Obviously,  the quantity $\sqrt{X}$ in this picture is identified with the specific enthalpy $w$.
Equation (\ref{eq5}) with (\ref{eq3008}) and (\ref{eq2011}) coincides with that of
 Ref.\  \cite{babichev} derived in field theory with a general Lagrangian of the form
 $\mathcal{L}=\mathcal{L}(X,\theta)$.

\section{Effective sound speed with external pressure}
\label{external}

Consider a fluid with internal variables $p$, $\rho$, and $n$. Suppose we apply to the fluid an external pressure $p_{\rm ext}$
so that the total pressure is
\begin{equation}
P=p+p_{\rm ext}.
\label{eq5001}
\end{equation}
The speed of sound is still defined by 
\begin{equation}
c_{\rm s}^{2}= \left.\frac{\partial p}{\partial \rho} \right|_s ,
\label{eq5002}
\end{equation} 
but the thermodynamic TdS equation (\ref{eq2012}) must include the external pressure, i.e.,
\begin{equation}
dW=Tds+\frac{1}{n} dP ,
\label{eq5003}
\end{equation}
where
\begin{equation}
W=\frac{P+\rho}{n}=w+\frac{p_{\rm ext}}{n}.
\label{eq5004}
\end{equation}
Then the sound speed is given by 
\begin{equation}
c_{\rm s}^{2}= \left.\frac{\partial (P-p_{\rm ext})}{\partial \rho} \right|_s =
\frac{n}{W}\left.\frac{\partial W}{\partial n}\right|_s
-\frac{\partial p_{\rm ext}}{\partial \rho}.
\label{eq5009}
\end{equation}
For an isentropic process from (\ref{eq5003}) it follows 
\begin{equation}
dP=ndW , \quad\quad  d\rho=Wdn,
\label{eq5010}
\end{equation}
so by making use of 
\begin{equation}
\frac{\partial}{\partial n}=W\frac{\partial}{\partial \rho}
\label{eq5011}
\end{equation}
we find 
\begin{equation}
c_{\rm s}^{2}= 
\frac{n}{w+p_{\rm ext}/n}\left( \left.\frac{\partial w}{\partial n}\right|_s
-\frac{p_{\rm ext}}{n^2}\right) .
\label{eq5012}
\end{equation}
Now we make the following ansatz
\begin{equation}
p_{\rm ext} = \alpha (p+\rho) ,
\label{eq5005}
\end{equation}
where $\alpha=\alpha(z)$ will be determined by the requirement that  the speed of sound
is well defined as $z\rightarrow 0$.
With this ansatz we find a modified expression for the sound speed
\begin{equation}
\tilde{c}_{\rm s}^{2}= \frac{1}{1+\alpha}\left(
\frac{n}{w}\left. \frac{\partial w}{\partial n}\right|_s-\alpha\right)
=\frac{c_{\rm s}^2-\alpha}{1+\alpha}.
\label{eq5006}
\end{equation}


\begin{thebibliography}{99}
	
\bibitem{babichev}
E.~Babichev, V.~Mukhanov, and A.~Vikman, 
JHEP {\bf 0802}, 101 (2008)
[arXiv:0708.0561 [hep-th]].
\bibitem{novello2} 
M.~Novello and E.~Goulart, 
Class.\ Quant.\ Grav.\  {\bf 28}, 145022 (2011)
[arXiv:1102.1913 [gr-qc]]. 
\bibitem{novello}
M.~Novello, E.~Bittencourt, U.~Moschella, E.~Goulart, J.~M.~Salim, and J.~D.~Toniato, 
JCAP {\bf 1306}, 014 (2013)
[arXiv:1212.0770 l[gr-qc]].
\bibitem{jacobson}
T.~Jacobson,
PoS \textbf{QG-PH}, 020 (2007)
[arXiv:0801.1547 [gr-qc]].	
%
\bibitem{visser}
M.~Visser, 
Class.\ Quant.\ Grav.\  {\bf 15}, 1767 (1998)
[arXiv:gr-qc/9712010].
\bibitem{bilic}
N.~Bili\'c, 
Class.\ Quant.\ Grav.\  {\bf 16},  3953 (1999)
[arXiv:gr-qc/9908002].
\bibitem{kinoshita} 
S.~Kinoshita, Y.~Sendouda, and K.~Takahashi, 
Phys.\ Rev.\ D {\bf 70}, 123006 (2004).
[astro-ph/0405149].
\bibitem{barcelo2} 
C.~Barcelo, S.~Liberati and M.~Visser,
Living Rev.\ Rel.\  {\bf 8}, 12 (2005)
[Living Rev.\ Rel.\  {\bf 14}, 3 (2011)]
[gr-qc/0505065].  
%
\bibitem{barbero1}
J.~F.~Barbero G.,
Phys.\ Rev.\ D {\bf 54}, 1492 (1996)
[arXiv:gr-qc/9605066].
\bibitem{barbero2}
J.~F.~Barbero G. and E.~J.~S.~Villasenor, 
Phys.\ Rev.\ D {\bf 68}, 087501 (2003)
[gr-qc/0307066].
\bibitem{mukohyama}
S.~Mukohyama and J.~P.~Uzan, 
Phys.\ Rev.\ D {\bf 87}, 065020 (2013)
[arXiv:1301.1361 [hep-th]].
%
	
	
	
\bibitem{hartnoll1}
S.~A.~Hartnoll, C.~P.~Herzog and G.~T.~Horowitz,
Phys. Rev. Lett. \textbf{101}, 031601 (2008)
[arXiv:0803.3295 [hep-th]].
\bibitem{hartnoll2}
S.~A.~Hartnoll, C.~P.~Herzog and G.~T.~Horowitz,
JHEP \textbf{12}, 015 (2008)
[arXiv:0810.1563 [hep-th]].
%
\bibitem{gubser1}
S.~S.~Gubser, C.~P.~Herzog, S.~S.~Pufu and T.~Tesileanu,
Phys. Rev. Lett. \textbf{103}, 141601 (2009)
[arXiv:0907.3510 [hep-th]].
%
\bibitem{gauntlett}
J.~P.~Gauntlett, J.~Sonner and T.~Wiseman,
Phys. Rev. Lett. \textbf{103}, 151601 (2009)
[arXiv:0907.3796 [hep-th]].
%
\bibitem{benini}
F.~Benini, C.~P.~Herzog, R.~Rahman and A.~Yarom,
JHEP \textbf{11}, 137 (2010)
[arXiv:1007.1981 [hep-th]].
\bibitem{horowitz1}
G.~T.~Horowitz,
Lect. Notes Phys. \textbf{828}, 313-347 (2011)
[arXiv:1002.1722 [hep-th]].
%
\bibitem{aprile}
F.~Aprile, D.~Roest and J.~G.~Russo,
JHEP \textbf{06}, 040 (2011)
[arXiv:1104.4473 [hep-th]].

\bibitem{bobev} 
  N.~Bobev, A.~Kundu, K.~Pilch and N.~P.~Warner,
  JHEP {\bf 1203}, 064 (2012)
  [arXiv:1110.3454 [hep-th]].
%

  \bibitem{albash} 
  T.~Albash and C.~V.~Johnson,
  JHEP {\bf 1205}, 079 (2012)
  [arXiv:1202.2605 [hep-th]].
%
 \bibitem{chakraborty} 
A.~Chakraborty,
Class. Quant. Grav. \textbf{37}, no.6, 065021 (2020)
doi:10.1088/1361-6382/ab6d09
[arXiv:1903.00613 [hep-th]].
%
  \bibitem{cai} 
  R.~G.~Cai, L.~Li, L.~F.~Li and R.~Q.~Yang,
  Sci.\ China Phys.\ Mech.\ Astron.\  {\bf 58}, no. 6, 060401 (2015)
  [arXiv:1502.00437 [hep-th]].
%
  \bibitem{hartnoll} 
  S.~A.~Hartnoll,
  Class.\ Quant.\ Grav.\  {\bf 26}, 224002 (2009)
  [arXiv:0903.3246 [hep-th]].  
%
\bibitem{hossenfelder} 
S.~Hossenfelder,
Phys.\ Lett.\ B {\bf 752}, 13 (2016)
[arXiv:1508.00732 [gr-qc]].
%
\bibitem{hossenfelder2} 
S.~Hossenfelder,
Phys.\ Rev.\ D {\bf 91}, no. 12, 124064 (2015)
[arXiv:1412.4220 [gr-qc]].

\bibitem{nikolic} 
N.~Bili\'c and H.~Nikolic,
Class.\ Quant.\ Grav.\  {\bf 35}, no. 13, 135008 (2018)
[arXiv:1802.03267 [gr-qc]].

\bibitem{zingg} 
N.~Bili\'c and T.~Zingg,
arXiv:1903.03401 [gr-qc]. 

%
\bibitem{fischbacher}
T.~Fischbacher, K.~Pilch and N.~P.~Warner,
[arXiv:1010.4910 [hep-th]]. 
%
  
\bibitem{randall1}
L.~Randall and R.~Sundrum, Phys.\ Rev.\ Lett. {\bf 83}, 3370 (1999)
%
\bibitem{randall2}
L.~Randall and R.~Sundrum, Phys.\ Rev.\ Lett. {\bf 83}, 4690 (1999)
%
%
\bibitem{bombelli}
L.~Bombelli, R.~K.~Koul, J.~H.~Lee and R.~D.~Sorkin,
Phys.\ Rev.\ D {\bf 34}, 373 (1986).

\bibitem{srednicki}
M.~Srednicki,
Phys.\ Rev.\ Lett.\  {\bf 71}, 666 (1993)
[arXiv:hep-th/9303048].
%
\bibitem{maldacena}
J.~M.~Maldacena,
Adv. Theor. Math. Phys. \textbf{2}, 231-252 (1998)
[arXiv:hep-th/9711200 [hep-th]].  
%
\bibitem{ryu}
  S.~Ryu and T.~Takayanagi,
  Phys. Rev. Lett. \textbf{96}, 181602 (2006)
  [arXiv:hep-th/0603001 [hep-th]];
  JHEP \textbf{08}, 045 (2006)
  [arXiv:hep-th/0605073 [hep-th]].
  \bibitem{lewkowycz}
  A.~Lewkowycz and J.~Maldacena,
  JHEP \textbf{08} (2013), 090
  [arXiv:1304.4926 [hep-th]].
  \bibitem{engelhardt}
  N.~Engelhardt and A.~C.~Wall,
  JHEP \textbf{01} (2015), 073
  [arXiv:1408.3203 [hep-th]].
  
\bibitem{pethick}  
C.~J.~Pethick and H.~Smith, {\it Bose-Einstein Condensation
in Diluted Gases}, Cambrige University Press, Cambrige
(2006).
\bibitem{uhlmann}
M.~Uhlmann, Y.~Xu and R.~Schutzhold,
New J. Phys. \textbf{7}, 248 (2005)
[arXiv:quant-ph/0509063 [quant-ph]].
\bibitem{girelli}
F.~Girelli, S.~Liberati and L.~Sindoni,
Phys. Rev. D \textbf{78}, 084013 (2008)
[arXiv:0807.4910 [gr-qc]].
\bibitem{fleurov}
V.~Fleurov and R.~Schilling. 
Phys.\ Rev.\ A 85, 045602 (2012)
[arXiv:1105.0799[cond-mat.quant-gas]].
\bibitem{rinaldi}
M.~Rinaldi,
Phys. Rev. D \textbf{84}, 124009 (2011)
[arXiv:1106.4764 [gr-qc]].
\bibitem{anderson}
P.~R.~Anderson, R.~Balbinot, A.~Fabbri and R.~Parentani,
Phys. Rev. D \textbf{87}, no.12, 124018 (2013)
[arXiv:1301.2081 [gr-qc]].
\bibitem{bilic2}
N.~Bilic and D.~Tolic,
Phys. Rev. D \textbf{87}, no.4, 044033 (2013)
[arXiv:1210.3824 [gr-qc]].
  \bibitem{bilic3}
N.~Bili\'c and D.~Toli\'c,
Phys. Rev. D \textbf{88}, 105002 (2013)
[arXiv:1309.2833 [gr-qc]].

  \bibitem{landau}
L.~D.~Landau, E.~M.~Lifshitz,
{\em Fluid Mechanics},
(Pergamon, Oxford, 1993) p. 507.
%


\bibitem{piattella} 
  O.~F.~Piattella, J.~C.~Fabris, and N.~Bili\'c, 
  Class.\ Quant.\ Grav.\  {\bf 31}, 055006 (2014)
  [arXiv:1309.4282 [gr-qc]].



\end{thebibliography}
\end{document}